\magnification1200
\rightline{KCL-MTH-00-33}
\rightline{CERN-TH/2000-147}
\rightline{hep-th/0005270}

\vskip .5cm
\centerline
{\bf{{ Hidden Superconformal Symmetry in M Theory}}}
\vskip 1cm
\centerline{ P. West}
\vskip .2cm
\centerline{Department of Mathematics}
\centerline{King's College, London, UK}
\centerline{ and }
\centerline{Theory Division }
\centerline{CERN}
\centerline{CH1211, Geneva 23}

\leftline{\sl Abstract}
The bosonic sectors of the eleven dimensional and IIA supergravity 
theories are derived as non-linear realisations. 
The underlying group includes the conformal 
group, the general linear group and as well as 
automorphisms of the supersymmetry algebra. 
We discuss the supersymmetric extension and 
argue that Osp(1/64) is a symmetry of M theory. 
We also derive the effective action of the closed bosonic 
string as a non-linear realisation. 
\vskip .5cm

\vfill\eject


\parskip=18pt

\medskip
{\bf {1.  Introduction}}
\medskip

Much of the progress in recent years  in our understanding of the 
non-perturbative 
effects of string theory has relied on the structure of the supergravity 
theories in 
eleven and ten dimensions. While there is only one    eleven dimensional 
supergravity theory [1],  in ten dimensions there exist 
 the IIA [2,3, 4] and  IIB [5,6,7] supergravity theories as well as 
 the type I 
supergravity theory coupled to the 
Yang-Mills theory [8] . These theories are essentially uniquely determined
by the type of supersymmetry that they possess.  Hence given a string theory 
in ten dimensions its 
complete low energy effective action must be the supergravity with 
the same space-time supersymmetry. 
One intriguing feature of supergravity theories is the occurance of 
coset space symmetries that  control the way the scalars in these theories 
behave. The four dimensional $N=4$ supergravity theory possess a  
SL(2,R)/ U(1) symmetry [9],  the IIA theory a  SO(1,1)  symmetry [2], 
the IIB 
theory a SL(2,R)/ U(1) [5] and the further 
reductions of the eleven dimensional 
 supergravity theory possess cosets based on the exceptional groups  
[10,11]. These symmetries have also played an 
important role in string dualities in recent years [12,13] and any further 
elucidation of the symmetries of  supergravity theories could 
prove useful. 
\par
Although these symmetries can be viewed as a consequence of supersymmetry, 
it is desirable to have a deeper understanding.   
One step in this direction has been the 
extension of the coset space description of  the scalars to include 
the gauge fields [14]. 
This has been achieved by introducing a group with 
Grassmann odd as well as  
Grassmann even generators.  All of these generators are 
scalars under the Lorentz group and  
the indices of the gauge fields are incorporated by writing them as 
forms. The group elements $g$ of the coset are then 
exponentials of these forms each of which is multiplied  
by  a generator  with the corresponding 
Grassmann character. This has the advantage that one automatically finds the 
gauge field strengths when taking the Cartan forms,  $g^{-1} d g$ 
 using the method of non-linear 
realisations [15,16]. The result is an elegant 
formulation of these sectors 
of the 
supergravity theories, but it is not apparent  how this method 
can be extended to 
include the graviton or indeed the fermionic sectors of the theory. 
\par
Recently, it was shown [17] that part of the $GL(32)$ automorphism 
group  of the 
supersymmetry algebra was found to be a symmetry of the fivebrane equations 
of motion. This symmetry was also found to play an important role in 
formulating the branes of M theory in terms of a non-linear realisation, 
indeed the world-volume gauge field strengths are the Goldstone bosons 
for part of this automorphism symmetry [18]. It was conjectured [17] that this 
symmetry could play a role in M theory and should occur in 
eleven dimensional supergravity. 
\par
Long ago [19],  Ogievetsky realised that the group of general 
coordinate realisations was the closure of the conformal group 
and the group of affine transformations, IGL(4) in four dimensions. 
 As a consequence, in reference [20], gravity was 
reformulated as the non-linear 
realisation of these two groups. Subsequently, it was shown [21] that the 
Sokatchev-Ogievetsky superspace formulation of the $N=1$ supergravity [22]  
in four dimensions could be expressed as a non-linear realisation
\par
In this paper we wish to revive this old idea of 
realising gravity as a non-linear realisation and,  by  combining it 
 with the presence of the automorphism symmetries,  show that the bosonic 
sector of  
eleven dimensional supergravity can be expressed as a non-linear realisation. 
In particular,  in section two, we will show how 
the bosonic sector of sector of eleven-dimensional supergravity, 
that is the graviton and the rank three gauge field,  is a 
non-linear realisation of the conformal group, SO(2,11) and a group 
which is generated by the generators of the group 
of affine transformations IGL(11) and two further generators which are of 
  rank three and rank six. While the graviton is the Goldstone boson for the 
group GL(11), the gauge field and 
its dual are the Goldstone bosons associated with these two additional 
generators. We will argue that these new generators arise as 
part of the GL(32) 
automorphism group of the supersymmetry algebra  
in the fully supersymmetric theory. 
One puzzle with realising gauge fields as Goldstone fields is that 
one does not obviously find their field strengths when following the 
standard method of non-linear realisations. In fact, 
the field strengths of these gauge fields arise only 
as a result of demanding that the theory be invariant under both groups. 
\par
We will also show,  in section three, that the bosonic sector of 
IIA supergravity can also be derived as a non-linear realisation. In section 
four,  we show that if one starts with a generic theory of Goldstone fields, 
some of which carry  anti-symmetrised space-time indices,  
and also demands conformal symmetry then one finds that 
these Goldstone fields must possess gauge symmetries.  
In section five,  we show that  the low energy effective action 
for the closed bosonic string can be written 
as a non-linear realisation. 
We explain,  in section six,  
how one may derive the dynamics of branes in a background 
using the theory of non-linear realisations. 
We sketch, in section seven,   how the 
non-linear realisation of the bosonic sectors of supergravity theories 
can be extended to the fully theory 
including the fermions. We conclude  in section eight.

\medskip
{\bf {2.  Eleven Dimensional Supergravity}}
\medskip
The Lagrangian of eleven dimensional supergravity written  in the signature 
$diag (\eta_{ab}) =(-1,1,1,\ldots  1)$ is given by [1] 
$$
L=+{e\over4\kappa^2}R\big(\Omega(e,\psi)\big)-
{e\over 48}F_{\mu_1\dots\mu_4}F^{\mu
_1\dots\mu_4}-{e\over2}
\bar\psi_\mu\Gamma^{\mu\nu\varrho}D_\nu\left({1\over2}
(\hat{\hat \Omega} +\hat\Omega)\right)\psi_\varrho
$$
$$
-{1\over 192}e\kappa
(\bar\psi_{\mu_1}\Gamma^{\mu_1\dots\mu_6}
\psi_{\mu_2}
+12\bar\psi^{\mu_3}
\Gamma^{\mu_4\mu_5}\psi^{\mu_6})(F_{\mu_3\dots\mu_6}+\hat
F_{\mu_3\dots\mu_6} )
$$
$$+{2 \kappa\over (12)^4}
\epsilon^{\mu_1\dots\mu_{11}}F_{\mu_1\dots\mu_4}F_{\mu_5
\dots\mu_8}A_{\mu_9\mu_{10}\mu_{11}}
\eqno(2.1)$$
where
$$ F_{\mu_1\dots\mu_4}=4\partial_{[\mu_1}A_{\mu_2\mu_3\mu_4]}
$$
$$
\hat F_{\mu_1\dots\mu_4}=F_{\mu_1\dots\mu_4}+3\bar\psi_{[\mu_1}
\Gamma_{\mu_2\mu_3}\psi_{\mu_4]},
\eqno(2.2)$$
and
$$\hat{\hat\Omega} _{\mu bc}= \hat \Omega _{\mu bc}+{1\over 4}
\bar \psi_\nu\Gamma_{\mu bc }^{\ \ \ \ \ \nu\lambda}\psi_{\lambda},
$$
$$\hat  \Omega _{\mu bc}= w _{\mu bc} (e)
-{1\over 2}
(\bar \psi_\nu\Gamma_{ c}\psi_{b}-\bar \psi_\nu\Gamma_{ b}\psi_{c}
+\bar \psi_c\Gamma_{ \mu}\psi_{b})
$$
and 
$$w _{\mu bc} (e) = {1\over 2}(e_b{}^\rho\partial_\mu e_{\rho c }
-e_c{}^\rho\partial_\mu e_{\rho b }) 
-{1\over 2}(e_b{}^\rho\partial _\rho e_{\mu c }-
e_c{}^\rho\partial _\rho e_{\mu b })
-{1\over 2} ( e_b{}^\lambda e_c{}^\rho\partial _\lambda e_{\rho a }  
 - e_c{}^\lambda e_b{}^\rho\partial _\lambda e_{\rho a })e_\mu{}^a
\eqno(2.3)$$
The symbol $ w _{\mu mn} (e) $ is the usual expression for the
spin-connection in terms of the vielbein $e_\mu^n$.  
\par
The equation of motion of the gauge field is given by 
$$
D _{\mu_1}F^{\mu_1\ldots \mu_4} + {(det e)^{-1}\over 4.(12)^2}
\epsilon ^{\mu_2\ldots \mu_4\nu_1\ldots \nu_8}F_{\nu_1\nu_2\nu_3\nu_4}
F_{\nu_5\nu_6\nu_7\nu_8}=0
\eqno(2.4)$$
This equation may be rewritten in first order form as 
$$
F_{\mu_1\ldots \mu_4}={det e\over 7!}
\epsilon _{\mu_1\ldots \mu_4\nu_1\ldots\nu_7} F^{\nu_1\ldots \nu_7}
\eqno(2.5)$$
where 
$$
\tilde F_{\nu_1\ldots \nu_7}\equiv 7(\partial_{[\nu_1}A_{\nu_2\ldots \nu_7]}
+5 A_{[\nu_1\ldots \nu_3}F_{\nu_4\ldots \mu_7]})
\eqno(2.6)$$
\par
We consider the Lie algebra whose non-vanishing  commutators are 
$$
[K^a{}_b,K^c{}_d]=\delta _b^c K^a{}_d - \delta _d^a K^c{}_b, \ 
[K^a{}_b,P_c]= - \delta _c^a P_b
$$
$$
[K^a{}_b, R^{c_1\ldots c_6}]= \delta _b^{c_1}R^{ac_2\ldots c_6}+\dots, \  
 [K^a{}_b, R^{c_1\ldots c_3}]= \delta _b^{c_1}R^{a c_2 c_3}+\dots,
\eqno(2.7)$$
$$  [ R^{c_1\ldots c_3}, R^{c_4\ldots c_6}]= 2 R^{c_1\ldots c_6} 
\eqno(2.8)$$
where $+\dots $ denote the appropriate anti-symmetrisations.  The generators 
$K^a{}_b$ and $P_c$ generate the affine group IGL(11) while  
the generators 
$R^{c_1\ldots c_3}$ and $ R^{c_1\ldots c_6}$ form a subalgebra that 
is the same as that which was found to be a symmetry of the fivebrane [17]. 
This subalgebra was also 
required in the description of the fivebrane as a non-linear realisation [18]. 
In these references it was identified as part of the GL(32) 
automorphism algebra of the eleven dimensional supersymmetry algebra.   
We denote  by $G_{11}$ the group whose   Lie algebra is that of 
equations (2.7) and (2.8). 
\par
We now construct the non-linear realisation corresponding to the group 
$G_{11}$ 
taking the Lorentz group to be a local symmetry.  
The generators of the 
Lorentz group are given by $J_{ab}=K_{ab}-K_{ba}$ where the 
indices are lowered with the Minkowski metric. We therefore consider group 
elements of the form  
$$g = e^{x^\mu P_\mu} e ^{h_{a}{}^b K^a{}_b} exp {({A_{c_1\ldots c_3} 
R^{c_1\ldots c_3}\over 3!}+ {A_{c_1\ldots c_6} R^{c_1\ldots c_6}\over 6!})} 
\eqno(2.9)$$
The fields  $h^a{}_b$, $A_{c_1\ldots c_3}$ and $A_{c_1\ldots c_6}$ 
depend on $x^\mu$. Although we use the exponential parameterization 
the reader who prefers a globally valid expression can readily 
rewrite the above group element in the appropriate form. In fact,  one could 
take $x^\mu$, $h^a{}_b$, $A_{c_1\ldots c_3}$ and $A_{c_1\ldots c_6}$ 
to depend on D parameters, thus leading to a kind of democracy 
between fields and coordinates. 
To recover the above form from this formulation, 
one uses the reparameterisation invariance inherent 
in the construction to choose the $x^\mu$ equal to the parameters. 
\par
The theory is to be invariant under 
$$g \to g_0 g h^{-1}
\eqno(2.10)$$
where $g_0$ is a rigid element of the full group generated by the above 
Lie algebra 
and $h$ is a local element of the Lorentz group.  The corresponding 
$g_0$ invariant forms are given by 
$${\cal V }= g^{-1}d g -w
\eqno(2.11)$$
where $w\equiv {1\over 2} dx^\mu w_{\mu b}{}^c J^b{}_c$ is 
the Lorentz connection and so transforms as 
$$w \to h w h^{-1} +h d h^{-1}
\eqno(2.12)$$
As a result 
$${\cal V} \to h {\cal V } h^{-1}
\eqno(2.13)$$
\par
This approach differs from that of reference [20] where the  
Lorentz symmetry was a rigid symmetry and the field $h_a{}^b$ was symmetric. 
The advantage of the approach adopted here is that one finds directly the 
vielbein formulation of general relativity and so the identification of 
part of the theory with 
general relativity is readily apparent. 
\par
Evaluating $\cal V$ 
we find that 
$${\cal V}= dx^\mu(e_\mu{}^a  P_a + \Omega _a{}^b K^a{}_b
+ {1\over 3!}\tilde D_\mu A_{c_1\ldots c_3} R^{c_1\ldots c_3}
+{1\over 6!}\tilde D_\mu A_{c_1\ldots c_6} R^{c_1\ldots c_6})
\eqno(2.14)$$
where 
$$e_\mu{}^a \equiv (e^h)_\mu{}^a, \ \tilde D_\mu A_{c_1\ldots c_3}\equiv  
 \partial_{\mu} A_{c_1 c_2 c_3} 
+ ((e^{-1}\partial _\mu e)_{c_1}{}^{b}A_{b c_2 c_3}+ \dots ), 
$$
$$ 
\tilde  D_\mu A_{c_1\ldots c_6}\equiv  
\partial _\mu A_{c_1\ldots c_6}+ 
((e^{-1}\partial _\mu e)_{c_1}{}^{b}A_{b c_2 \ldots c_6}+ \dots )
- (A_{[ c_1\ldots c_3}\tilde D_\mu A_{c_4\ldots c_6]})
$$
$$\Omega_{\mu b}{}^c\equiv (e^{-1}\partial _\mu e)_b{}^c 
-w_{\mu b}{}^c,
\eqno(2.15)$$
where $+\ldots $ denotes the action of $(e^{-1}\partial _\mu e)$ on the other 
indices of $A_{c_1\ldots c_3}$  and $A_{c_1\ldots c_6}$. 
\par
The covariant derivatives of the Goldstone fields 
 associated with this non-linear realisation, that is of the field $h_a{}^b$ 
and the fields $A_{c_1\ldots c_3}$  and $A_{c_1\ldots c_6}$,  are given by 
$$\Omega_{a b}{}^c\equiv (e^{-1})_a{}^\mu (e^{-1}\partial _\mu e)_b{}^c 
-w_{a b}{}^c,
$$
$$ 
\tilde D_a A_{c_1\ldots c_3} 
\equiv (e^{-1})_a{}^\mu \tilde D_\mu A _{c_1\ldots c_3},\ 
\tilde D_a A_{c_1\ldots c_6} \equiv (e^{-1})_a{}^\mu 
\tilde D _\mu A_{c_1\ldots c_6}
\eqno(2.16)$$
We note that these quantities are not  field strengths 
as the indices are not anti-symmetrised,  nor are the derivatives those 
that occur in general relativity. As we shall see,  we will only recover the 
field strengths of the equations of motion once we consider the simultaneous 
non-linear realisation with the conformal group. 
\par
Under 
$h = e^{({1\over 2}w_a{}^b(x) J^a{}_b)}$,  we find that $e_\mu{}^a$ 
transforms as a elfbein should under a local Lorentz transformation and 
indeed all the indices of the fields in ${\cal V}$ 
which are contracted with the generators 
 are rotated in the correct way as to be interpreted as tangent space indices. 
 A matter 
field $B$ transforms as $B \to B'= D(h)B$ where $D$ is the representation of 
the Lorentz group to which $B$ belongs. 
The covariant derivative of the  matter 
field $B$ 
is given by 
$$\tilde D_a B \equiv (e^{-1})_a{}^\mu \partial _\mu B +
{1\over 2} w_{a b}{\ }^c \Sigma ^b{}_c B ,
\eqno(2.17)$$
where $w_{a b}{\ }^c\equiv(e^{-1})_a{}^\mu  w_{\mu b}{\ }^c$ 
and   
$\Sigma_{a}{}^b$ is the representation of the generators of the Lorentz group  
associated with $B$. 
\par
As we have mentioned above,  the bosonic sector of 
eleven dimensional supergravity 
is not uniquely determined by just taking a 
non-linear realisation of the group $G_{11}$. We must find a 
non-linear realisation of the closure of this group with the conformal group. 
We could calculate the closure and find the non-linear realisation 
of this infinite dimensional group. However, it is easier to find 
the simultaneous realisation of the two groups, that is calculate 
the Cartan forms for both groups and then use  only those combinations that 
are invariant under both groups. We must also take into account any 
duplication of Goldstone fields between the two groups. We will now 
follow this second stratergy. 
\par
We now construct the 
non-linear realisation for the conformal group SO(2,11), 
 taking the now rigid Lorentz group as the istropy group. This procedure 
is well known, [20,23], but for completeness we summarise the derivation. 
The generators of the conformal group obey the relations 
$$[J_{ab}, K_c]= -\eta_{ac}K_b +\eta_{bc}K_a, 
[P_a, D]= P_a, \ [K_a, D]= -K_a, \ [P_a, K_b]= +2\eta _{ab}D -2J_{ab}
\eqno(2.18)$$ 
in addition to those of the Poincare group and relations where the 
commutators vanish. 
We take as our 
coset representative 
$$g = e^{x^\mu P_\mu}e^{\phi^\mu K_\mu}e^{\sigma D}
\eqno(2.19)$$
and the Cartan forms are given by 
$$g^{-1}d g= dx^a(e^\sigma P_a+ e^{-\sigma}(\partial_a \phi^b - 
\phi^c\phi_c \delta _a^b +2 \phi_a \phi ^b)K_b
+(\partial _a \sigma +2 \phi_a)D 
+ (-\delta _a^c \phi^d+\delta _a^d \phi^c)J_{cd}) 
\eqno(2.20)$$
The covariant derivatives of the Goldstone 
fields are obtained by taking all the above 
expressions, with the exception of the first term, 
 and multiplying by $e^{-\sigma}$. 
These  transform only under the Lorentz group 
and we can set the covariant derivative 
 for $\sigma$ to zero and still preserve the group. 
As a result,  we can eliminate $\phi_\mu$ in terms of $\partial_\mu \sigma$, 
namely $2 \phi_\mu =-\partial_\mu \sigma$. 
In effect, this  
leaves  $\sigma $ as the only Goldstone field. 
\par
It is straightforward to find the transformations  
under dilations and special conformal transformations
of $\sigma$ and a field 
$B$  which transforms under the representation $\Sigma _{ab}$ of the 
Lorentz group. 
The result is 
$$\delta \sigma = (2x\cdot \beta x\cdot \partial -x^2 
\beta\cdot \partial )\sigma +2 \beta_\mu  x^\mu+ \lambda, \ 
\delta B= (2x\cdot \beta x\cdot \partial -x^2 
\beta\cdot \partial )B+(\beta^a x^b- \beta^b x^a) \Sigma _{ab} B
\eqno(2.21)$$
No dilation term occurs in the above variation since dilations are not 
part of the isotropy group. 
\par
The covariant derivative with respect to conformal transformations,
 denoted $\Delta_a$,  of the  field $B$  
is given by 
$$\Delta_a B = e^{-\sigma}(\partial_a + \partial^b \sigma \Sigma_{ab})B
\eqno(2.22)$$
In particular, for a vector field $A_a$ we have 
$$\Delta_a A_b= e^{-\sigma}(\partial _a A_b +\eta_{ab}\partial^c \sigma A_c
-\partial _b \sigma A_a)
\eqno(2.23)$$
\par
Following the procedure of 
Borisov and Ogievetski [20], we must now construct quantities from the 
derivatives of the Goldstone 
fields of  the first group $G_{11}$  which are also 
covariant with respect to  the conformal group. In view of the identical 
transformation of 
$x^\mu$ under dilations and the  determinant part of GL(11) we must identify 
$h_\mu{}^a= \bar h_\mu^a+ \sigma\delta_\mu^a$ where $\bar h_\mu{}^\mu=0$. 
While the field $\sigma$ identified in this way must transform in the 
relavent way determined by the conformal group, 
the fields $\bar h_\mu{}^a$, $A_{c_1\ldots c_3}$ and 
$A_{c_1\ldots c_6}$ transform under conformal transformations as their 
indices suggest. It is simplest to first carry out this procedure for the 
fields $A_{c_1\ldots c_3}$  and $A_{c_1\ldots c_6}$ . 
The $G_{11}$ covariant derivative of $A_{c_1\ldots c_3}$ can be rewritten as 
$$\tilde D_a A_{c_1\ldots c_3}= (\bar e)_a{}^\mu(\Delta _\mu A_{c_1\ldots c_3}
  +e^{-\sigma}(-\eta _{\mu c_1}\partial^d \sigma A_{d\ldots c_3} 
$$
$$+D_{c_1}\sigma A_{\mu\ldots c_3}
+ \partial _\mu \sigma A_{c_1\ldots c_3}
+(\bar e^{-1}\partial_\mu \bar e)_{c_1}{}^d   A_{d \ldots c_3}+\ldots))
\eqno(2.24)$$
In this equation $\bar e = e^{\bar h}$  and $+\ldots $ denotes the 
terms that arise when the connections of the derivatives 
contract with the other indices on $A_{c_1\ldots c_3}$. 
Even at order $(\bar h)^0$ it is apparent that only by completely 
anti-symmetrising in $a, c_1 \ldots c_3$ can one obtain an expression 
such that all $\sigma$ dependence is through 
 the conformal derivative $\Delta _a$ 
alone and as a result 
 is an expression that  is simultaneously covariant under both 
groups. Thus the unique simultaneously covariant quantity 
is 
$$\tilde F_{c_1\ldots c_4}
\equiv  4(e_{[ c_1}{}^\mu \partial _\mu A_{c_2\ldots c_4]}
+ e_{[ c_1}{}^\mu ( e^{-1} \partial_\mu  e)_{[ c_2 }{\ \  }^b 
A_{b c_3 c_4]}+\ldots)
\eqno(2.25)$$
 A similar calculation for the gauge field $A_{c_1\ldots c_6}$ 
leads to the unique simultaneously covariant expression 
$$
\tilde F_{c_1\ldots c_7}\equiv  
7(e_{[ c_1}{}^\mu(\partial _\mu A_{c_2\ldots c_7]})
+ e_{[ c_1}{}^\mu ( e^{-1} \partial_\mu  e)_{[ c_2 }{\ \  }^b 
A_{b c_3\ldots c_7]}+\ldots  +5 \tilde F_{[c_1\ldots c_4}\tilde
F_{c_5\ldots c_7]})  
\eqno(2.26)$$
\par 
What is not apparent from the above expressions is that the covariant 
derivatives are those that one should find in general relativity. To 
verify this, and indeed to recover general relativity itself,  
we must  
recover the usual expression for the spin connection in terms of the elfbein. 
We can use the inverse Higgs effect [24] to place constraints on the 
covariant derivative $\Omega_{ab}{\ }^c$ of the field $h_\mu{}^a$.  
Within the context of the group $G_{11}$ there is no unique way to do this. 
However, as explained above, 
we must do this in just such a way that the $G_{11}$ covariant 
derivative of equation (2.17) can be rewritten in terms of the 
covariant derivatives of the conformal group of equation (2.21). 
Consider a matter field $B$ as discussed above, its $G_{11}$ 
covariant derivative  can be expressed as
$$\tilde D_a B= (\bar e)_a{}^\mu (\Delta _\mu B 
- e^{-\sigma} \partial_\nu\sigma  \Sigma _\mu{}^\nu B 
+{1\over 2}e^{-\sigma} w_{\mu b}{\ }^c \Sigma ^b{}_c B)
\eqno(2.27)$$ 
This equation tells us that  $w_{ab}{\ }^c$ must be expressible as 
in terms 
of the conformal covariant derivatives of $\bar e_\mu{}^a$ 
as well as a derivatives of $\sigma$ which must cancel  
 the second term in the right-hand side of 
the above equation. The unique solution is to take the constraint 
$$\Omega_{a[bc]}- \Omega_{b(ac)}+ \Omega_{c(ab)}=0 
\eqno(2.28)$$
This results in the well known expression for the spin connection 
in terms of the elfbein given in equation (2.3). Although the connection 
term that appears in the field strengths of equations (2.25) and (2.26) 
looks incorrect,  when one takes into account
the anti-symmetry on all the indices one finds that the covariant derivative 
can be re-expressed 
in terms of the usual spin connection appropriate to tangent indices. 
Thus the field strengths of these equations  
when written in terms of the components appropriate 
for the coordinates of the space-time are 
just the curl of the gauge potential written in the same components. 
\par
The equations of motion for the simultaneous non-linear realisation 
must be written in terms of $\tilde F_{c_1\ldots c_4}$ 
and $ \tilde F_{c_1\ldots c_7}$ 
given  in equations (2.25) and (2.26) and the 
spin-connection in such a way that 
the equations are covariant under the local Lorentz transformations. 
Clearly, the spin connection can only enter either in   
$\tilde F_{c_1\ldots c_4}$ and $ \tilde F_{c_1\ldots c_7}$,  in the way 
which is already specified,  or through the Riemann tensor 
$$R_{\mu\nu b}{\ \ }^c\equiv \partial _\mu w_{\nu b}{\ }^c
+w_{\mu b}{\ }^d w_{\nu d}{\ }^c -(\mu \to \nu) 
\eqno(2.29)$$
\par
The unique first order equation for the gauge field which is not trivial 
is 
$$  \tilde F_{c_1\ldots c_4} = {1\over 7!} \epsilon _{c_1\ldots c_{11}}
\tilde F^{c_5\ldots c_{11}}
 \eqno(2.30)$$
in agreement with equation (2.5) when written in the local coordinate 
frame. The only other non-trivial equation is  
$$ R_{\mu\nu b}{\ \ }^c e_c{}^\nu e_a{}^\mu -{1\over 2} \eta_{ab} 
R_{\mu\nu b}{\ \ }^c e_c{}^\nu e^{b\mu}-{c\over 4} 
(\tilde F_{a c_1\ldots c_3}
\tilde F_{b}{}^ {c_1\ldots c_3}-{1\over 6}\eta_{ab}
\tilde F_{c_1\ldots c_4}
\tilde F^{c_1\ldots c_4})=0
\eqno(2.31)$$
as it should  be. The constant $c$, of proportionality 
can only be determined once the 
full supersymmetric treatment is given. It has value 1. 
\medskip 
{\bf {3. Gauge Symmetry}}
\medskip
 It is instructive to trace more precisely how the gauge 
invariance of the fields $A_{a_1\ldots a_3}$ and $A_{a_1\ldots a_6}$ 
 arises as a consequence of the 
simultaneous realisation of $G_{11}$ and the conformal group. Taking 
$$g_0=exp({c_{\mu_1\ldots \mu_3}\delta^{\mu_1\ldots \mu_3}_{a_1\ldots a_3}  
R^{a_1\ldots a_3}\over 3!}+
 {c_{\mu_1\ldots \mu_6}\delta^{\mu_1\ldots \mu_6}_{a_1\ldots a_6} 
R^{a_1\ldots a_6} \over 6!})
\eqno(3.1)$$ 
where 
$ \delta^{\mu_1\ldots \mu_n}_{a_1\ldots a_n}= \delta _{\mu_1}^{a_1}\ldots
\delta _{\mu_n}^{a_n}$  and 
$ c_{\mu_1\ldots \mu_3}$ and  $c_{\mu_1\ldots \mu_6}$  are constants 
in equation (2.10),  we find that 
the vielbein is inert and the other fields transform as 
$$\delta A _{a_1\ldots a_3}= c_{a_1\ldots a_3}, \ 
\delta A_{a_1\ldots a_6}= c_{a_1\ldots a_6}+ 20
c_{[a_1\ldots a_3}A_{a_4\ldots a_6]}
\eqno(3.2)$$
where $c_{\mu_1\ldots \mu_3}= e_{\mu_1}{}^{a_1}\ldots e_{\mu_3}{}^{a_3}
c_{a_1\ldots a_3}$ and similarly for $c_{a_1\ldots a_6}$. 
The vielbeins occur because  
the factor in $g$ which contains the   fields $A _{a_1\ldots a_3}$ and 
$A _{a_1\ldots a_6}$  
is to the right 
 of that 
containing the $h_a{}^b$  fields. Thus,  it is the   fields with 
curved indices that transform most simply. 
To find the conformal transformation 
of the fields with curved indices we write them as 
$A_{\mu_1\ldots \mu_p}= (e^{\bar h})_{\mu_1}{}^{a_1}\ldots (e^{\bar h})
_{\mu_p}{}^{a_p}e^{p\sigma}A_{a_1\ldots a_p}$ for $p=3,6$. 
Taking into account the 
conformal transformation of $\sigma$ and $B$ of equation (2.21) 
we find that 
$$\delta A_{\mu_1\ldots \mu_p}= (2(x\cdot \beta)( x\cdot \partial) -x^2 
(\beta\cdot \partial) )A_{\mu_1\ldots \mu_p} +
(2\beta_{\mu_1} x^\kappa A_{\kappa\mu_2\ldots \mu_p}
-2x_{\mu_1}\beta^\kappa A_{\kappa\mu_2\ldots \mu_p}+\dots)
$$
$$+2p(\beta\cdot x)A_{\mu_1\ldots \mu_p}
\eqno(3.3)$$
where $+\ldots $ stands for the other terms where the induced 
Lorentz rotation 
acts on the other indices of $ A_{\mu_1\ldots \mu_p}$. We note that this 
is the variation of a matter field that we would have 
obtain had we included the 
dilations in the isotropy group and assigned the field dilation 
weight p.  
\par
For simplicity,  we will illustrate the mechanism of how gauge 
symmetry arises for a single form  field 
$A_{\mu_1\ldots \mu_p}$ that has a constant shift, i.e.  
$ \delta A_{\mu_1\ldots \mu_p}=  c_{\mu_1\ldots \mu_p}$,  
under a non-linear realisation. Carrying out the commutation of this 
shift with a special conformal transformation 
we find that 
$$[\delta _c, \delta _\beta] A_{\mu_1\ldots \mu_p}= 
p \partial _{[\mu_1} \tilde \Lambda_{\mu_2\ldots \mu_p]}^{(2)}
\eqno(3.4)$$ 
We recognise this as a gauge transformation 
with parameter 
$$\tilde \Lambda _{\mu_2\ldots \mu_p}^{(2)}=2 p x\cdot \beta x^\kappa 
c_{\kappa \mu_2\ldots \mu_p}-x^2 \beta^\kappa c_{\kappa \mu_2\ldots \mu_p}
+(-2x_{\mu_2}\beta^\kappa x^\rho c_{\rho \kappa \mu_3\ldots \mu_p} +\dots ) 
\eqno(3.5)$$ 
We may write the original shift of the field as 
a gauge transformation with parameter 
$  \Lambda ^{(1)}_{\mu_2\ldots \mu_p} = x^\kappa 
c_{\kappa \mu_2\ldots \mu_p}$ and taking its  commutator with 
 special conformal transformations we find another gauge transformation 
which is quadratic in $x^\mu$. Thus starting from a gauge transformation 
that is linear in $x^\mu$ we obtain one which is bilinear. 
\par
By induction, we will now show  that   
taking repeated commutators with special 
conformal transformations leads to a gauge transformation with an 
arbitrary local parameter. Let us suppose that after taking $r-1$ 
commutators we find a transformation which can be written as a 
gauge transformation of 
$A_{\mu_1\ldots \mu_p}$, denoted  
$\Lambda_{\mu_2\ldots \mu_p}^{(r)}$, which is a polynomial in 
$x^\mu$ of degree r. Taking the 
commutator of this with another special conformal transformation we find that 
 $$[\delta _\Lambda,\delta _\beta ]A_{\mu_1\ldots \mu_p}=
p \partial _{[\mu_1} \tilde \Lambda_{\mu_2\ldots \mu_p]}^{(r+1)}
\eqno(3.6)$$ 
where 
$$\tilde \Lambda_{\mu_2\ldots \mu_p]}^{(r+1)}
 = (2x\cdot \beta x\cdot \partial -x^2 
\beta\cdot \partial )\Lambda_{\mu_2\ldots \mu_p}^{(r)} 
$$
$$+
(2\beta_{\mu_2} x^\kappa \Lambda _{\kappa\mu_3\ldots \mu_p}^{(r)}
-2x_{\mu_2}\beta^\kappa \Lambda_{\kappa\mu_3\ldots \mu_p}^{(r)}+\dots)
+2(p-1)x\cdot \beta \Lambda_{\mu_2\ldots \mu_p}^{(r)}
\eqno(3.7)$$
Hence we recover another gauge transformation, which is a 
polynomial in $x^\mu$ of one degree higher. It is clear that 
proceeding in this way we can find an arbitrary local gauge transformation. 
Thus,  we have shown that taking 
the closure of a Goldstone shift and the conformal 
group leads to a local gauge transformation. In fact, if we start from a 
non-linear realisation of the fields  we can regard gauge 
invariance as a consequence of conformal invariance. 
\par
The standard $U(1)$ field has been been previously considered as a 
Goldstone boson by considering an infinite dimensional algebra [29].  
\medskip
{\bf {4.  IIA Supergravity}}
\medskip
The bosonic part of the ten-dimensional IIA  supergravity theory is given by
  [2],[3],[4]
$$\eqalign{
L^B&=eR\big(w(e)\big)-{1\over12}ee^{\sigma\over2}
F_{\mu_1\dots\mu_4}^\prime F^
{\prime \mu_1\dots\mu_4}-{1\over3}ee^{-\sigma}F_{\mu_1\dots\mu_3}
F^{\mu_1\dots
\mu_3}\cr
&\quad-ee^{{3\over2}\sigma}F_{\mu_1\mu_2}F^{\mu_1\mu_2}
-{1\over2}\partial_\mu
\sigma\partial^\mu\sigma\cr
&\quad+{1\over2\cdot(12)^2}\epsilon^{\mu_1\dots\mu_{10}}
F_{\mu_1\dots\mu_4}F_{
\mu_5\dots\mu_8}A_{\mu_9\mu_{10}}\cr}
\eqno(4.1)$$
where
$$F_{\mu_1\mu_2}=2\partial_{[\mu_1}A_{\mu_2]}
\eqno(4.2)$$
$$
F_{\mu_1\mu_2\mu_3}=3\partial_{[\mu_1}A_{\mu_2\mu_3]}
\eqno(4.3)$$
$$
F_{\mu_1\dots\mu_4}^\prime =4(\partial_{[\mu_1}A_{\dots\mu_4]}
+2A_{[\mu_1}F_{\mu_2\mu_3\mu_4]})
\eqno(4.4)$$
\par
The equations of motion of the bosonic, non-gravitational sector of the theory 
were expressed as a non-linear realisation in reference [14]. Although we use 
a different group  and strategy, 
some of the steps below have analogues 
in those of reference [14]. 
We proceed much as for eleven-dimensional supergravity, 
we take the group $G_{IIA}$ to have the generators of 
IGL(10)   together with the generators 
$R_{a_1\ldots a_p}$ for $p=0,1,2,3,5,6,7,8$ 
that obey equations  analoguous to  
equation (2.7 ) as well as  the relations 
$$ [R,R^{a_1\ldots a_p}]=c_p R^{a_1\ldots a_p} , \ 
[R^{a_1\ldots a_p},R^{a_1\ldots a_q}]= c_{p,q} R^{a_1\ldots a_{(p+q)}}
\eqno(4.5)$$
where 
$$c_{1}=-c_{7}=-{3\over 4},\ c_{2}=-c_{6}={1\over 2},\ 
 c_{3}=-c_{5}=-{1\over 4}:
$$
$$  c_{1,2}=-c_{2,3}=-c_{3,3}=c_{2,5}= c_{1,5}=2,\ 
c_{1,7}=3, \ c_{2,6}=2,\ c_{3,5}=1
\eqno(4.6)$$
and all  other $c$'s vanish. We take the group element of $G_{IIA}$
 to be of the form 
$$g= e^{x^\mu P_\mu}g_hg_A
\eqno(4.6)$$
where 
$$
g_h= e^{h_a{}^b K^a{}_b},
$$ and 
$$ g_A= exp ({A_{a_1\ldots a_8}\over 8!}
R^{a_1\ldots a_8})
$$
$$\ldots exp ({A_{a_1\ldots a_3}\over 3!}
R^{a_1a_2a_3}) 
exp ({A_{a_1 a_2}\over 2!}
R^{a_1 a_2})exp (A_a R^a)exp (A R)
\eqno(4.7)$$
We also take the Lorentz group to be a local symmetry and so consider the 
quantity 
$${\cal V}= g^{-1}d g-w 
\eqno(4.8)$$
We may rewrite this as 
$${\cal V}= (g_h^{-1}dg_h)+
(g_A^{-1}d g_A +g_A^{-1}(g_h^{-1}d g_h)g_A -g_h^{-1}dg_h)
$$
$$\equiv dx^\mu(e_\mu{}^a P_a+\Omega _{\mu a}{\ }^b K^a{}_b)
+dx^\mu (\sum _{p=1}^{8}{1\over p!}
\tilde D_\mu A_{a_1\ldots a_p}R^{a_1\ldots a_p})
\eqno(4.9)$$ 
where the definition applies to each of the terms in the brackets separately. 
\par
The next step is to 
demand conformal invariance. In particular,  we should take combinations of 
derivatives of the 
Goldstone fields that are also conformally covariant. The procedure 
follows  closely 
those of sections two and three. Indeed, 
it is an inevitable consequence of section three that 
 the fields $A_{a_1\ldots a_p}$ will appear in quantities that are 
gauge invariant. 
  Thus the quantities which involve the fields $A_{a_1\ldots a_p}$ that 
are $G_{IIA}$ 
and conformally covariant are 
$$ \tilde F_{a_1\ldots a_p}\equiv 
p e^{-c_{(p-1)}A}\tilde D _{[a_1}A_{a_2\ldots a_3]}
\eqno(4.10)$$
\par
One finds  that 
$$\tilde F_a= D_a A, \ 
\tilde F_{a_1a_2}= 2 e^{{3\over 4}A}D_{[a_1}A_{a_2]}, \ 
\tilde F_{a_1a_2a_3}= 3 e^{-{1\over 2}A}D_{[a_1}A_{a_2a_3]},  
$$
$$
\tilde F_{a_1a_2a_3 a_4 }= 4 e^{{1\over 4}A}(D_{[a_1}A_{a_2a_3a_4]}
+2 e^{{1\over 2}A}A_{[a_1}\tilde F_{a_2a_3a_4]}), 
\tilde F_{a_1a_2a_3 a_4 a_5}=0,
$$
$$ 
\tilde F_{a_1\ldots  a_6 }= 
6 e^{-{1\over 4}A}( D_{[a_1}A_{a_2\ldots a_6]}
+5 e^{-{1\over 4}A} A_{[a_1a_2}\tilde F_{a_3\ldots a_6]}), 
$$
$$
\tilde F_{a_1\ldots  a_7 }= 7  e^{{1\over 2}A}(D_{[a_1}A_{a_2\ldots a_7]}
-20  A_{[a_1a_2a_3} D_{[a_4}A_{a_5\ldots a_7]}
+2 e^{{1\over 4}A}A_{[a_1}\tilde F_{a_2\ldots a_7]}), 
$$
$$
\tilde F_{a_1\ldots  a_8}= 8  e^{-{3\over 4}A}
(D_{[a_1}A_{a_2\ldots a_8]}
-7.6 A_{[a_1a_2}(D_{a_3}A_{a_4\ldots a_8]}
+10 A_{[a_3a_4}(D_{[a_5}A_{a_6\ldots a_8]})),  
$$
$$
\tilde F_{a_1\ldots  a_9}= 9(D_{[a_1}A_{a_2\ldots a_9]}
-8.7 A_{[a_1a_2}D_{[a_3}A_{a_4\ldots a_9]}  +
{7.4\over 3} e^{{1\over 4}A}A_{[a_1a_2a_3}\tilde F_{a_4\ldots a_9]}
+3e^{{3\over 4}A} A_{[a_1}\tilde F_{a_2\ldots a_9]})
\eqno(4.11)$$
\par
The unique equations of motion of the forms that are first order in 
derivatives and constructed from the simultaneously 
 covariant  derivatives of 
the Goldstone fields are 
$$
\tilde F^{a_1\ldots a_p}= {1\over (10-p)!}\epsilon ^{a_1\ldots a_{10}}
 \tilde F_{a_{(p+1)}\ldots a_{10}},\ p=1,2,3,4
\eqno(4.12)$$
as well as the equation for the vielbein. 
These are the equations of motion of IIA supergravity. 
\par
One might have thought that the easiest way to obtain the 
non-linear realisation of IIA supergravity 
would be to directly 
perform the dimensional reduction on the non-linear realisation of 
eleven dimensional supergravity. However, 
the correct number of fields  and generators does not arise in a natural way 
as becomes  
apparent if one tries. 
This suggests that the   formulation of eleven dimensional supergravity 
given in section two may not be the most natural one and that  there should 
exist  a first order formulation of the vielbein equation of motion 
by introducing higher rank fields. 
\medskip
{\bf {5. The Closed Bosonic String Effective Action }}
\medskip
One can also apply the theory of non-linear realisations to the 
low energy effective action of the closed string. This has been found to be 
[26] 
$$\int d^D x det e(R-{4\over (D-2)}\partial _\mu \phi\partial ^\mu \phi
-{1\over 3} e^{-{8\phi\over D-2}}F_{\mu_1\mu_2\mu_3}F^{\mu_1\mu_2\mu_3})
\eqno(5.1)$$
where $F_{\mu_1\mu_2\mu_3}= 3\partial_{[\mu_1}A_{\mu_2\mu_3]}$. 
In this action, $D$ is the dimension of space-time, but we must  take $D=26$ 
to obtain the consistent closed bosonic string. In principle we should include 
the cosmological term for $D\not= 26$ and there will also be 
corrections to the dilaton potential from higher order genus surfaces [26]. 
Presumeably,  these terms could be accounted for by 
taking into account an anomaly in 
the symmetry 
below.  
\par
We then consider the group $G_{D}$ whose generators 
are $K^a{}_b$, $R$,  $R^{a_1a_2}$, $R^{a_1\ldots a_{(D-4)}}$ 
and $R^{a_1\ldots a_{(D-4)}}$. They obey equations analogous to 
equation  (2.7)as well as the relations
$$ [R,R^{a_1\ldots a_p}]=c_p R^{a_1\ldots a_p} , \ 
[R^{a_1\ldots a_p},R^{a_1\ldots a_q}]= c_{p,q} R^{a_1\ldots a_{(p+q)}}
\eqno(5.2)$$
where 
$$c_{2}=-c_{D-4}={4\over (D-2)},\ 
 c_{2,D-4}=2. 
\eqno(5.3)$$
We take 
and all the other $c$'s vanish. 
\par
The non-linear realisation of $G_{D}$ is built out of the group element 
$g= g_h g_A$ where 
$$g_A= exp ({A_{a_1\ldots a_{(D-2)}}R^{a_1\ldots a_{(D-2)}}\over (D-2)!})
exp ({A_{a_1\ldots a_{(D-4)}}R^{a_1\ldots a_{(D-4)}}\over (D-4)!})
exp ({A_{a_1a_2}R^{a_1a_2}\over (2)!})exp (AR)
\eqno(5.4)$$
Calculating the Cartan forms $g^{-1}dg-w$ and demanding simultaneous 
invariance under the conformal group,  we find that the equations of motion 
must be built out of $w_{ab}^{\ c}$ and 
$$\tilde F_a=D_a A, \ 
\tilde F_{a_1a_2a_3}=3e^{-{4\over (D-2)}A}D_{[a_1} A_{a_2a_3]},\ 
$$
$$
\tilde F_{a_1\dots a_{(D-3)}}=(D-3)e^{{4\over (D-2)}A}D_{[a_1} 
A_{a_2\dots a_{(D-3)}]},
$$
$$ 
\tilde F_{a_1\dots a_{(D-1)}}=(D-1)(D_{[a_1} A_{a_2\dots a_{(D-1)}]}
+(D-2)e^{-{4\over (D-2)}A}A_{[a_1a_2}F_{a_3\dots a_{(D-1)]}})
\eqno(5.5)$$
The equations of motion are given by 
$$\tilde F^{a_1\ldots a_p}= {1\over (D-p)!}
\epsilon^{a_1\ldots a_D}\tilde F_{a_{(p+1)}\ldots a_{D}}, \ p=1,2
\eqno(5.6)$$
provided we identify $\phi$ with $A$, as well as the vielbein equation. 
\medskip
{\bf {6. Branes in a Background}}
\medskip  
In a recent paper [18],  the branes of M theory 
were derived as a non-linear realisation. Since in this paper we have shown 
that the background supergravity to which they couple can also be formulated  
as a  non-linear realisation, it is straightforward, at least 
in principle, to  describe the dynamics of branes in a background as a 
non-linear realisation. We now 
illustrate the procedure  for the case of a bosonic brane coupled to gravity. 
\par
We begin with a  group element of IGL(D)  
of the form 
$$g= e^{X^{\underline a}(\xi) P_{\underline a}}
e^{h_{\underline a}{}^{\underline b}(X)K^{\underline a}{}_{\underline b}} 
\eqno(6.1)$$
We use the same index notation as in reference [18], 
where $D$ is the dimension of space-time and 
$\xi^n$ are 
the coordinates of the brane worldvolume. We consider the forms 
$${\cal V}= g^{-1}d g -w\equiv d\xi^n(e_n{}^a P_a+ f_n{}^{a'}P_{a'}+ 
\omega_{n\underline a}{\ }^{\underline b} K^{\underline a}{}_{\underline b})
\eqno(6.2)$$
The spin connection $w$ takes values in the Lie algebra of 
$SO(1,p)\times SO(D-p-1)$. 
\par
In fact, in reference [18],  we took the isotropy 
group to be SO(1,p), although  we could have taken the above group. 
Making this latter choice simplifies the analysis of reference [18] 
 a bit, but the results are the same. 
\par
Returning to the bosonic brane in a background, 
we find  that 
$$e_n{}^a= \partial _n X^{\underline m} e_{\underline m}{}^a, \ 
f_n{}^{a'}= \partial _nX^{\underline m} e_{\underline m}{}^{a'}
\eqno(6.3)$$
Since $f_n^{a'}$ transforms in a covariant manner,  we can set it to zero. 
This solves for $\partial_n X^{a'}$ in terms of the $e_n{}^{a'}$ 
of the background gravity. Hence, in the case of  a  
local background we find that the parts of the vielbein that belong to the 
coset 
$SO(1,D-1)\over SO(1,p)\times SO(D-p-1)$ play the role of the Goldstone bosons 
of the Lorentz group that were solved for in reference [18]. Proceeding 
as in that paper,  and using the constraint $f_n{}^{a'}=0$, we find that 
$$e_n{}^a\eta_{ab}e_m{}^b= \partial_n X^{\underline p}
g_{\underline p\underline  q} 
\partial_m X^{\underline q}
\eqno(6.4)$$
where $g_{\underline n\underline  m} = e_{\underline m}{}^{\underline a}
\eta _{\underline a\underline  b}
e_{\underline n}{}^{\underline b}$. Thus, we find that the invariant action is 
$$\int d^p \xi det e_n^a= \int d^p \xi\sqrt{-det (\partial_n X^{\underline p}
g_{\underline p\underline  q}\partial_m X^{\underline q})}
\eqno(6.5)$$
as it should be. 
\medskip
{\bf {7. Supersymmetric Extension}}
\medskip 
It would be interesting to extend the analysis of this paper to the 
full supergravity theories, that is incorporate supersymmetry. Let us 
first sketch 
how this would go for eleven dimensional supergravity. 
To extend the group IGL(11),  it is natural to consider the group IGL(11/32). 
The generators of GL(11/32) group can be labelled by 
$K^A{}_B$ where $A= (a,\alpha)$
and similarly for $B$ etc. We can then denote the generators by 
$K^A{}_B=(K^a{}_b, K^a{}_\alpha, K^\alpha{}_a, K^\alpha{}_\beta)$ 
and the generators of inhomogeneous transformations by $P_a, K_\alpha$. 
The non-linear realisation is then built from the group elements of the form 
$$g= e^{(X^aP_a+K^\alpha \theta_\alpha)}e^{(h_a{}^bK^a{}_b+A_\alpha{}^\beta 
K^\alpha{}_\beta)}
e^{(\psi_a{}^\alpha K^a{}_\alpha+ \zeta_\alpha{}^a K^\alpha{}_a)}
\eqno(7.1)$$
 \par
In the group element of equation (7.1),   
the fields $h_a{}^b$, $A_\alpha{}^\beta$, $\psi_a{}^\alpha$ and 
$\zeta_\alpha{}^a$ are functions of $x^a$ and $\theta_\alpha$ and 
have geometric dimensions 0, 0, 1/2 and -1/2 respectively. Thus the 
lowest components of $h_a{}^b$, $A_\alpha{}^\beta$ and $\psi_a{}^\alpha$
have the correct dimensions to be identified with the graviton, 
gauge fields and gravitino respectively.  Indeed, 
their shifts under the appropriate 
symmetries of the non-linear realisations make this identification 
inevitable. 
\par
We must also consider a non-linear realisation of a 
supersymmetric generalisation of the conformal group. It is 
known  [25], 
that there is a unique generalisation of SO(2,11) that also contains the 
supersymmetry algebra: it is $Osp(1/64)$. The Lie algebra of this group 
can be written as 
$$[R_{\hat \alpha \hat \beta},R_{\hat \gamma \hat \delta}]= 
-C_{\hat \beta  \hat \gamma} R_{\hat \alpha \hat \delta}
-C_{\hat \alpha  \hat \delta } R_{\hat \gamma \hat \beta}-
C_{\hat \beta  \hat \delta} R_{\hat \alpha \hat \gamma}
-C_{\hat \alpha  \hat \gamma} R_{\hat \beta \hat \delta}\ 
$$
$$
\{ \rho_{\hat \alpha}, \rho_{\hat \beta} \}= R_{\hat \alpha \hat \beta}, \ 
[\rho_{\hat \gamma},R_{\hat \alpha \hat \beta}]=
C_{\hat \beta  \hat \gamma}\rho_{\hat \alpha}
+C_{\hat \alpha \hat \gamma}\rho_{\hat \beta}
\eqno(7.2)$$
where $C_{\hat \beta  \hat \gamma}=-C_{\hat \gamma  \hat \beta}$ 
is the metric that occurs in the invariant line element of this group and 
$\hat \alpha, \hat \beta=1,2\ldots ,64$. 
\par
We  now  decompose the 
64 component spinor, $\rho_{\hat \alpha}$ in this group into two 32 component 
spinors using the index decomposition $\hat \alpha = (\alpha, \alpha')$ 
where $\alpha =1,\ldots ,32, \alpha' =1,\ldots , 32$ etc. 
In particular, we set 
 $Q_\alpha= \rho_{ \alpha}$ and 
$S^\alpha= \rho_{\beta'}C^{\beta' \alpha}$, $R_{\alpha\beta}= Z_{\alpha\beta}$, 
$R_{\alpha}{}^\beta=R_{\alpha\beta'}C^{\beta' \beta}$ and  
$Z^{\alpha\beta}= R_{\alpha'\beta'}C^{\alpha' \alpha}C^{\beta' \beta}$. 
Taking $C_{\alpha\beta}=0=C_{\alpha'\beta'}$ we may write the 
algebra of Osp(1/64) in the form 
$$\{Q_\alpha, Q_\beta\}= Z_{\alpha \beta}, \  
[Q_\alpha,Z_{\gamma \delta}]=0,\  [Z_{\alpha \delta},Z_{\gamma \beta}]=0,
$$
$$[Q_\alpha, R_\gamma{}^\delta]= -\delta_\alpha^\delta Q_\gamma,\ 
[Z_{\alpha \beta},R_{\gamma}{}^{ \delta}]=
-\delta_\alpha^\delta Z_{\gamma \beta}
-\delta_\beta^\delta Z_{\gamma \alpha}
\eqno(7.3)$$
and 
$$
\{S^\alpha, S^\beta\}= Z^{\alpha \beta}, \ 
[S^\alpha,Z^{\gamma \delta}]=0,\  
[Z^{\alpha \delta},Z^{\gamma \beta}]=0,
$$
$$
[S^\gamma, R_\alpha{}^\beta]= \delta_\alpha^\gamma S^\beta,\ 
[Z^{\alpha \beta},R^{\gamma}{}_{ \delta}]=
\delta_\gamma^\beta Z^{\alpha \delta}
+\delta^\alpha_\gamma Z^{\delta \beta}
\eqno(7.4)$$
as well as 
$$\{Q_\alpha, S^\beta\}= R_{\alpha}{}^ \beta, \ 
[Z_{\alpha\beta}, Z^{\gamma\delta}]=-\delta_\beta^\gamma R_\alpha{}^\delta
-\delta_\alpha^\delta R_\beta{}^\gamma-\delta_\beta^\delta R_\alpha{}^\gamma -
\delta_\alpha^\gamma R_\beta{}^\delta
\eqno(7.5)$$
We recognise that Osp(1/64) contains a sub-algebra, given in equation (7.3),  
which is precisely the usual supersymmetry algebra in eleven dimensions with 
all its central charges, plus the  GL(32) 
automorphism group 
that was found to play a role in the fivebrane equations of motion [17] and 
in the branes of M theory realised as a non-linear realisation [18]. 
Expanding in $\gamma$-matrices, we can express 
$$R^\alpha{}_\beta= \sum_n R^{a_1\ldots a_n}
(\gamma_{a_1\ldots a_n})^\alpha{}_\beta .
\eqno(7.6)$$ 
The  generators 
$R$ and $R_{a_1a_2}$ are to be identified with dilations and Lorentz 
rotations. 
\par
The considerations of this paper show that Osp(1/64) must be a 
symmetry of eleven dimensional supergravity. 
This group has previously been considered [30] in the context of M theory 
with two times and mentioned as a possible unifying group in reference [31]. 
\par
When taking the simultaneous realisation of the two groups 
IGL(11/32)  and Osp(1/32) 
we must identify the Goldstone fields whose corresponding generators 
have the same action on the coordinates $x^a$ and $\theta^\alpha$. 
In principle, 
 one should also consider the action of Osp(1/64) on 
the central charges, but for the 
present discussion we shall ignore this subtilty. As for the 
bosonic sector consider in section two,  
the dilations are in common and so we must  
identify  $h^\mu{}_\mu$ with $\sigma$,  However,  the 
generators $K^\alpha{}_\beta$ and $R_\alpha{}^\beta$ act the same way 
on the coordinates $x^a$ and $\theta^\alpha$ with the exception of the 
scalar and rank two generators that behave differently.   In  
Osp(1/64) these are the dilations and Lorentz rotations and 
so their actions on the coordinates $x^a$ and $\theta^\alpha$ 
are related.   In contrast,  the GL(11/32) 
action on the coordinates $x^a$ and $\theta^\alpha$  is unrelated. 
Thus, we should 
identify all of the generators in 
$K^\alpha{}_\beta$ with those in 
$R_\alpha{}^\beta$ with the exception of the two generators  
of rank zero and two. As a result, in the simultaneous non-linear realisation 
we find  
that we have generators of every rank in $R_\alpha{}^\beta$,  including the 
dilations and Lorentz rotations, as well as two additional generators of 
  rank zero and two, which we may denote by $K$ and $K_{ab}$. 
Hence, in eleven dimensions,   
we find  the automorphisms 
$R^{a_1\ldots a_n}$ for $n=0,1,2,3,4,6$ 
and the two additional  
generators, $K$ and $K_{a_1a_2}$. 
It was observed on reference [17] that 
the algebra of equation (2.8) was a contraction of the GL(32) 
automorphism algebra. Hence, it would seem natural to identify the 
generators  $R^{a_1\ldots a_3}$ and $R^{a_1\ldots a_6}$,  which are the 
generators,  whose Goldstone fields are 
the gauge fields of eleven dimensional supergravity in the 
non-linear realisation of section two,  with 
the automorphisms that arise in the groups GL(11/32) and Osp(1/64). 
\par
The supergravity action of equation (2.1) essentially contains 
three contributions, 
the kinetic terms in the first line, the Noether term in the second line and 
the Chern-Simmons term in the last line. We have already accounted for the 
first and last terms and in the supersymmetric extension we must 
account for the 
 Noether term. In the Cartan forms of  the group element of equation (7.1)
we find a term of the form
$$e^{-\psi_a{}^\alpha K^a{}_\alpha}
(\tilde D_a A_{a_1\ldots a_3}R^{a_1\ldots a_3}+
\tilde D_a A_{a_1\ldots a_6}R^{a_1\ldots a_6}
)e^{\psi_a{}^\alpha K^a{}_\alpha}
\eqno(7.7)$$
Taking the commutator of the $R^{a_1\ldots a_3}  $ and 
$R^{a_1\ldots a_6} $ generators with $K^a{}_\alpha$ to be the obvious 
$\gamma$-matrix times  $K^a{}_\alpha$ we do indeed find a term that 
has, at least in form, that of the Noether term. 
\par
We now briefly comment on the supersymmetric extension of the   IIA 
theory considered in section four. 
The extension of the group IGL(10) is presumably the group IGL(10/32). 
Although reference [25] was concerned with $N=1$ supersymmetry, it 
would seem inevitable that the unique 
extension of the conformal group to include a 
type II superalgebra is the group Osp(1/64). We should consider 
the simultaneous realisation of both of these groups. 
In the later group, we will find the GL(32) automorphism group and so 
the generators $R^{a_1\ldots a_p}  $ for $p=0,1\ldots 10$. 
Identifying the generators in the same way as above we find that 
the simultaneous non-linear realisation of these two groups 
includes the generators 
$R^{a_1\ldots a_n}$ for $n=0,1,2, \ldots , 10$ 
as well as  two additional  
generators, which we can denote by $K$ and $K_{a_1a_2}$. We note that in 
contrast to the eleven dimensional theory most of the generators are 
need to ensure the necessary Goldstone bosons. The correspondence 
with the automorphisms of the supersymmetry algebra is less obvious 
in this case and it is possible that one may have to introduce generators in 
addition to those of the above two groups. In particular, the generator $R$,  
which leads to the SO(1,1) transformations of  the IIA supergravity theory, 
does not seem to have an obvious identification with these generators.  
\par
The non-linear realisation of the 
IIb supergravity theory follows a similar pattern, but 
the automorphisms that are active are different from those in the eleven 
dimensional and IIA supergravity theories. 
\par
The above is a sketch of the extension to the supergravity theory, 
however, until one actually carries out the full calculation one 
cannot be sure that all the considerations in this section are correct. 
\medskip
{\bf {8. Conclusion}}
\medskip
It is clear that all supergravity theories can be formulated as 
non-linear realisations. The bosonic part of the 
group underlying these constructions will 
include,  the conformal group, the general linear group and 
certain automorphisms. 
In the complete theory, the conformal group will be 
embedded in the  relevant 
Osp group which automatically contain the automorphisms of the Poincare 
supersymmetry algebras with all their central charges. 
\par
For many years it has 
been a puzzle to understand why the scalars that occur in 
supergravity theories 
belong to a non-linear realisation. However, from the perspective of this 
paper this it could be viewed as 
 just a consequence of the whole theory being a non-linear 
realisation. 
It is known that if one reduces eleven dimensional supergravity on a torus 
one finds [14] the group GL(11-d) in d dimensions. From the view point 
of the non-linear realisation of eleven dimensional 
supergravity given in this paper,  this is 
hardly surprising since it is just part of the original GL(11) 
group of the original theory. 
 However,  
it would be good to understand the emergence of the exceptional groups 
from the dimensional reduction  of the non-linear realisations given in 
this paper. 
\par
One intriguing feature of the constructions of this paper 
is that the group is apparently 
different for each supergravity theory.  
This would be compatible with the suggestion,  in references [17,18],  
that the full automorphism group is a symmetry of M theory and that 
as one goes to the limits of M theory such  as eleven 
dimensional supergravity, IIA and IIB theory one finds different 
contractions of this automorphism group.  In 
fact,  it is inevitable that Osp(1/64) is a symmetry of M theory 
as this group is the unique extension of the conformal group 
to include supersymmetry and is in required in  both 
the IIA and eleven dimensional 
supergravities. As 
the IIB
supersymmetry algebra can be otained form the IIA supersymmetry algebra 
by an invertible transformation [28], it is likely that 
Osp(1/64) is also required in the IIB case. 
It is interesting to note that by taking this group one 
automatically encodes  
 all the central charges and the GL(32) automorphism. 
As such, this group implicitly includes  
all the branes. 
Since it  includes brane rotating symmetries one would have to 
restrict the field of the group to ensure it was compatible with the charge 
quantization conditions.   
\par
The maximal supergravities in ten dimensions are the low energy 
limits of the coresponding string theories.  As such, 
 it is perhaps not surprising that they  should  possess a non-linear 
realisation in that this has been the traditional role for such formulations. 
However, it does suggest that there is an alternative formulation of 
these string theories, perhaps M theory,  
in which all the symmetries discussed in this paper, 
including Osp(1/64) are 
linearly realised. 
\par
>From a practical view point,  it would be interesting to see if one could use 
the conformal symmetry, and its superextension, to derive constraints on 
the Greens functions of the supergravity theories.  Such a calculation 
 would utilise our knowledge of solving conformal Ward identities with 
theorems about the behaviour of Greens functions of Goldstone particles. 
In a sense gravity and supergravity can be thought of as the analogues of the 
conformally invariant two dimensional models. One can think of 
the symmetry of these  latter models as being found by 
starting  with the finite dimensional globally defined 
conformal group and generating  an infinite dimensional group by taking its 
 closure with the group whose generators are   $L_2, L_0, L_2$. 
In the theories considered in this paper, 
 one also starts with a finite dimensional,  globally defined,  extension of 
conformal group 
and generates an infinite dimensional group 
by taking its closure with an extension of the affine group. 
\medskip
{\bf {Acknowledgment}}
\medskip
The author would like to thank Bernard Julia  for explaining the 
content of reference [14], Toine van Proeyen  for discussions on 
superconformal symmetry in eleven dimensions,  Arkardy Tseytlin 
for discussions on the closed bosonic string effective action 
  and George Papadopoulos 
for commenting on the manuscript. This work was
supported in part by the EU network  on Integrability,
Non-perturbative effects, and Symmetry in Quantum Field theory
(FMRX-CT96-0012).

\medskip
{\bf {References}}
\medskip
\parskip 0pt

\item{[1]} E. Cremmer, B. Julia and J. Scherk, Phys. Lett. 76B
(1978) 409.
\item{[{2}]} C. Campbell and P. West,
{\it ``$N=2$ $D=10$ nonchiral
supergravity and its spontaneous compactification.''}
Nucl.\ Phys.\ {\bf B243} (1984) 112.
\item{[{3}]} M. Huq and M. Namazie,
{\it ``Kaluza--Klein supergravity in ten dimensions''},
Class.\ Q.\ Grav.\ {\bf 2} (1985).
\item{[{4}]} F. Giani and M. Pernici,
{\it ``$N=2$ supergravity in ten dimensions''},
Phys.\ Rev.\ {\bf D30} (1984) 325.
\item{[5]} J, Schwarz and P. West,
{\it ``Symmetries and Transformation of Chiral
$N=2$ $D=10$ Supergravity''},
Phys. Lett. {\bf 126B} (1983) 301.
\item {[6]} P. Howe and P. West,
{\it ``The Complete $N=2$ $D=10$ Supergravity''},
Nucl.\ Phys.\ {\bf B238} (1984) 181.
\item {[7]} J. Schwarz,
{\it ``Covariant Field Equations of Chiral $N=2$
$D=10$ Supergravity''},
Nucl.\ Phys.\ {\bf B226} (1983) 269.
\item{[8]} L. Brink, J. Scherk and J.H. Schwarz, {\it
``Supersymmetric Yang-Mills Theories''},
Nucl. Phys. {\bf B121} (1977) 77;
F. Gliozzi, J. Scherk and D. Olive, {\it ``Supersymmetry,
Supergravity Theories and the Dual Spinor Model''}, Nucl. Phys. {\bf
B122} (1977) 253,  A.H. Chamseddine, {\it ``Interacting supergravity
in ten dimensions: the role of the six-index gauge field''},
Phys. Rev. {\bf D24} (1981) 3065;
E.\ Bergshoeff, M.\ de Roo, B.\ de Wit and P.\ van
Nieuwenhuizen, {\it ``Ten-dimensional Maxwell-Einstein
supergravity, its currents, and the issue of its auxiliary
fields''}, Nucl.\ Phys.\ {\bf B195} (1982) 97;
E.\ Bergshoeff, M.\ de Roo and B.\ de Wit, {\it ``Conformal
supergravity in ten dimensions''}, Nucl.\ Phys.\ {\bf B217} (1983)
143,  G. Chapline and N.S. Manton,
{\it ``Unification of Yang-Mills
theory and supergravity in ten dimensions''}, Phys. Lett. {\bf
120B} (1983) 105.  
{\item{[9]} S.\ Ferrara, J.\ Scherk and B.\ Zumino, 
``Algebraic Properties of Extended Supersymmetry''},
Nucl.\ Phys.\ {\bf B121} (1977) 393;
E.\ Cremmer, J.\ Scherk and S.\ Ferrara, {\it ``SU(4) Invariant
Supergravity Theory''}, Phys.\ Lett.\ {\bf 74B} (1978) 61.
\item{[10]} E. Cremmer and B. Julia,
{\it ``The $N=8$ supergravity theory. I. The Lagrangian''},
Phys.\ Lett.\ {\bf 80B} (1978) 48
\item{[11]} B.\ Julia, {\it ``Group Disintegrations''},
in {\it Superspace \&
Supergravity}, p.\ 331,  eds.\ S.W.\ Hawking  and M.\ Ro\v{c}ek,
Cambridge University Press (1981).
\item{[12]} A. Font, L. Ibanez, D. Lust and FD. Quevedo, Phys. Lett. B249
(1990) 35.
\item{[13]} C.M. Hull and P.K. Townsend,
{\it ``Unity of superstring  dualities''},
Nucl.\ Phys.\ {\bf B438} (1995) 109, hep-th/9410167.
\item{[14]} E. Cremmer, B. Julia, H. Lu and C. Pope, Dualisation of 
dualities II: Twisted self-duality of duobled fields and superdualities, 
hep-th/9806106 
\item{[15]} S. Coleman, J. Wess and B. Zumino, { Phys. Rev.\/}
{\bf 177} (1969) 2239;\
K. Callan, S. Coleman, J. Wess and B. Zumino,
{ Phys. Rev.\/} {\bf 177} (1969) 2247.
\item{[16]}D. V. Volkov, { Sov. J. Part. Nucl.\/} {\bf 4} (1973) 3;
D. V. Volkov and V. P. Akulov, { JETP Letters} {\bf 16} (1972) 438;
{ Phys. Lett. \/} {\bf B46} (1973) 109.
\item{[17]} O. Barwald and P. West, { Brane Rotating symmetries and
the fivebrane equations of motion} hep-th/9912226.
\item{[18]} P. West, { Automorphisms, Non-linear Realizations  and  Branes}, 
hep-th//0001216, JHEP to be published. 
\item{[19]} V. Ogievetsky, Lett. Infinte-dimensional algebra of general 
covariance group as the closure of the finite dimensional algebras 
of conformal and linear groups. 
Nuovo. Cimento, 8 (1973) 988.
\item{[20]} A. Borisov and V. Ogievetsky, 
Theory of dynamical affine and confomral 
symmetries as the theory of the gravitational field, 
Teor. Mat. Fiz. 21 (1974) 329. 
\item{[21]} E. Ivanov and J. Niederle, 
{$N=1$ supergravity as a non-linear realisation}, Phys. Rev D45, (1992) 4545. 
\item{[22]}V. Ogievetsky and E. Sokatchev, Phys. Lett. 79B (1978) 222. 
\item{[23]} C. Isham, A. Salam and J. Strathdee, Ann. Phys. A 16 (1983) 2571.
\item{[24]} E. Ivanov and V. Ogievetsky, Teor. Mat. Fiz. 25 (1975) 164. 
\item{[25]} J. van Holten and A. van Proeyen, $N=1$ supersymmetry algebras 
in $d=2,3,4$ mod 8, J. Phys A, 15 (1982) 376. 
\item{[26]} E. Fradkin and A. Tsetylin, Nucl Phys. B261 (1985) 1; 
Phys. Lett. B155 (1985) 316, 
\item {[27]}for a two reviews  see, 
Sigma Models  and Renormalization of string Loops, 
A.A. Tseytlin, Sigma Models  and Renormalization of string Loops, 
 
Lectures given at 1989 Trieste Spring School on Superstrings, Trieste, Italy, 
 Published in Trieste Superstrings 1989:0487-549;
  and 
String Effective Action; String Loop Corrections, 
Int.J.Mod.Phys.A3 (1988) 365-395.  
\item{[28]} E. Bergshoeff and A. van Proeyen, 
The Many Faces of Osp(1/32), hep-th/0003261.
\item{[29]}E. Ivanov and V. Ogievetsky, 
Gauge theories as theories of spontaneous breakdown, 
Lett. in Math. Phys. 1 (1976) 309. 
\item{[30]} Itzhak Bars, Cemsinan Deliduman and Djordje Minic, 
Phys.Lett.B457:275-284,1999: hep-th/9904063. 
\item{[31]} M. Gunaydin, Unitary supermulitplets of Osp(1/32,R) 
and M theory, hep-th/9803138
\end